

\magnification\magstephalf
\tolerance 10000

\font\rfont=cmr10 at 10 true pt
\def\ref#1{$^{\hbox{\rfont {[#1]}}}$}


\font\twelverm=cmr12
\font\twelvebf=cmbx12
\font\twelveit=cmti12

\def\0{\over } \def\1{\vec } \def\2{{1\over2}} \def\4{{1\over4}}
\def\5{\bar } 
\def\6{\partial }
\def\7#1{{#1}\llap{/}}
\def\8#1{{\textstyle{#1}}} \def\9#1{{\bf {#1}}}

\let\a=\alpha \let\b=\beta \let\g=\gamma \let\d=\delta
\let\e=\varepsilon   \let\th=\theta
 \let\k=\kappa  \let\m=\mu
\let\n=\nu  \let\p=\pi \let\r=\rho \let\s=\sigma
   
  \let\P=\Pi 
\let\L=\Lambda  \let\D=\Delta 
\def\mn{{\m\n}}

\def\({\left(} \def\){\right)} \def\<{\langle } \def\>{\rangle }
\def\[{\left[} \def\]{\right]}  

\let\el=\eqalign \let\en=\eqno

\def\a {\alpha} \def\b {\beta} \def\g {\gamma} \def \d {\delta}
\def\e{\epsilon}  
\def\s {\sigma}  \def\la{\lambda}
 \def\L {\Lambda} 
\def\pd {\partial}
\def\pmb#1{\setbox0=\hbox{#1}
 \kern.05em\copy0\kern-\wd0 \kern-.025em\raise.0433em\box0 }

\def \half {{\scriptstyle {1 \over 2}}}

\def \quarter {{\scriptstyle {1 \over 4}}}

 %

\def \i {\item}  

\def \cl {\centerline}
\parskip=6pt
\parindent=0pt
\hsize=17truecm\hoffset=-5truemm
\voffset=-0.5truecm\vsize=25.5truecm
\def\footnoterule{\kern-3pt
\hrule width 17truecm \kern 2.6pt}

\def\k{{\bf k}}

{\nopagenumbers
{\twelverm
\rightline{CERN-TH-6491/92}
\rightline{ENSLAPP-A-383/92}
\vfil
\centerline{{\twelvebf COVARIANT GAUGES AT FINITE TEMPERATURE}}
\bigskip
\bigskip
\cl{P V Landshoff}
\smallskip
\cl{{\twelveit
CERN, Geneva\footnote{$^{\dag}$}{On leave from DAMTP, University of
Cambridge}}}
\bigskip
\cl{A Rebhan{\twelveit\footnote{$^\ddagger$}{On leave from ITP,
Technical University of Vienna; boursier du
Minist\`ere des Affaires Etrang\`eres 
   }}}
\smallskip
\cl{{\twelveit Laboratoire de Physique Th\'eorique
ENSLAPP\footnote{$^*$}{URA 14-36
du CNRS, associ\'ee \`a l'ENS de Lyon, et au LAPP
d'Annecy-le-Vieux}, Annecy-le-Vieux}}
\vskip 3 truecm
{\twelvebf Abstract}
\bigskip
A  prescription is presented for real-time finite-temperature perturbation
theory in covariant gauges, in which only the two physical degrees of freedom
of the gauge-field propagator acquire thermal parts. The propagators for the
unphysical degrees of freedom of the gauge field, and for the
Faddeev-Popov ghost field,
are independent of temperature. This prescription is applied to the calculation
of the one-loop gluon self-energy and the two-loop interaction pressure, and
is found to be simpler to use than the conventional one.
\vfil
\line{ENSLAPP-A-383/92 \hfill}
\line{CERN-TH-6491/92  \hfill}

May 1992 (corrected June 1993)

\eject}}
\pageno=1

1. {\bf INTRODUCTION}
\medskip
In finite-temperature field theory, one calculates a grand partition
function
$$
Z=\sum _i \;\langle\; i\;|e^{-\b H}|\;i\;\rangle
\eqno(1.1a)
$$
and thermal averages
$$
\langle Q \rangle =Z^{-1}\sum _i \;\langle\; i\;|e^{-\b H}Q|\;i\;\rangle
.\eqno(1.1b)
$$
In both cases, the sum is over a complete orthonormal set of physical
states. In scalar field theory, these states span the whole Hilbert
space, and so one may replace the sums with traces of the operators.
But in the case of a gauge theory the Hilbert space contains also
unphysical states and so things are more complicated.

In this paper, we reconsider what to do about this for the case of
covariant gauges.
The traditional
solution\ref{1} is, in effect, to sum over all states of the gauge
particles, and then cancel the unphysical contributions with ghosts.
We shall argue that, for calculational purposes, it
may be simpler
to include in the sums (1.1) only the physical states from the start,
and not to introduce unphysical terms that have to be cancelled.

The finite-temperature propagator consists of a vacuum
piece, plus a ``thermal'' part. In real-time perturbation theory,
which we consider initially, it is a $2\times 2$ matrix\ref{2}.
Its vacuum piece is
$$
i{\bf D}^{\mu\nu}(x)=\left (\matrix{
\langle 0|TA^{\mu}(x)A^{\nu}(0)|0\rangle&
            \langle 0|A^{\mu}(- i\s ,{\bf 0})A^{\nu}(x)|0\rangle\cr
\langle 0|A^{\mu}(x^0- i\s ,{\bf x})A^{\nu}(0)|0\rangle&
                     \langle 0|\bar TA^{\mu}(x)A^{\nu}(0)|0\rangle\cr}
                              \right )
\eqno(1.2)
$$
with $\bar T$ the anti-time-ordering operator, {and
$0\le\s\le\b$.}
There is a similar matrix ghost-field propagator.
The elements of the thermal part of the matrix take account of all
the possible states of the heat bath and  involve the Bose
distribution.
In the traditional approach to covariant gauges, all the components
of the gauge-field propagator have a thermal part, as also does the ghost
propagator. We shall argue that it is simpler to give only the
propagators for the physical degrees of freedom a thermal part,
so that the unphysical components of the gauge field, and also the ghost
field, have only vacuum propagators. In the next section,
we give a derivation of this from the Gupta-Bleuler method of
quantising the gauge field, though in a sense the result is obvious
and can be written down without any long derivation.

This is because the thermal part of a propagator contains a $\d$-function
that puts the corresponding particle on shell. The thermal
part represents the absorption of a particle from the heat bath
or the emission of one into it. That is, the thermal part is directly
associated with the particles in the physical  states $|\;i\;\rangle$
summed in (1.1). Consequently 
only the physical degrees of freedom
acquire thermal parts for their propagation.

The thermal part of the propagator for the gauge field therefore has
the tensor structure
$$
\sum_{\la =1}^2 e^{\mu}_{\la}(k)e^{*\nu}_{\la}(k)
\eqno(1.3a)
$$
where $e_{\la}(k)$ are polarisation vectors and $k^2=0$.
There is some freedom in
the choice of these, but since the heat bath breaks the Lorentz
symmetry already,
the obvious choice is the one that causes no additional
Lorentz-symmetry breaking. So we choose polarisation vectors orthogonal
to the 4-velocity $u$ of the heat bath, and then (1.3a) is
$$
T^{\mu\nu}(k,u)=-g^{\mu\nu}+{{k^{\mu}u^{\nu}+u^{\mu}k^{\nu}}\over{u.k}}
        -u^2{{k^{\mu}k^{\nu}}\over{(u.k)^2}}
.\eqno(1.3b)
$$
In the frame where the heat bath is at rest this vanishes if either
$\mu$ or $\nu$ is zero, and its other elements are
$$
T^{ij}(k)=\d ^{ij} -{{k^ik^j}\over{{\k}^2}}
.\eqno(1.3c)
$$

To summarise what we have said, part of the finite-temperature
propagator of the gauge
field is the vacuum piece,
which in Feynman gauge reads
$$
-g^{\mu\nu}\left (\matrix{
{1/(k^2+i\e)}&
-2\pi i\;\delta ^{(-)}(k^2)\cr
-2\pi i\;\delta ^{(+)}(k^2)&{-1/(k^2-i\e)}\cr
                             }\right )
.\eqno(1.4a)
$$
We have not written explicitly the colour factor $\d ^{ab}$. We have made
the particular choice $\s =0$ in (1.2) so that the vacuum part of the
propagator is temperature-independent; for other choices of $\s$ it depends
on $\b$, but in a simple way.
To (1.4a) must be added the thermal piece of the propagator:
$$
-iT^\mn(k)\; 2\pi\d (k^2)\, n(|k_0|) 
\left (
    \matrix{1&1\cr1&1\cr}
                                   \right )
,\eqno(1.4b)
$$
where $n(x)=1/(e^{\b x}-1)$.

There is also the ghost propagator; this has only the vacuum
part, which is just equal to the matrix that appears within (1.4a).
We derive the propagator (1.4) in section 2 and generalise it to other gauges,
both covariant and noncovariant.

In order to show that this prescription can be
simpler than the conventional
one, we use it in section 3 to calculate the gluon self-energy and the
plasma pressure to lowest non-trivial order in the coupling.
Section 4 is a discussion
of our results.
In an Appendix we introduce some alternative versions of the
formalism in which the $2\times 2$ propagator matrix is diagonal
and temperature-independent, and the dependence on the temperature is
instead transferred to the interaction vertices.

\bigskip
2. {\bf GUPTA-BLEULER THEORY AT FINITE TEMPERATURE}
\medskip
To evaluate the grand partition function (1.1a) or a thermal
average (1.1b), we need to express the Hamiltonian in operator
form and to identify the physical states. For this, we use
standard Gupta-Bleuler theory\ref{3}.
We go through the details only for the case
of the Feynman gauge.

Start with the standard QCD Lagrangian $-\quarter F^2$,
with the addition of a gauge-fixing term but initially no ghost part:
$$
{\cal L}=-\quarter F^{\mu\nu}F_{\mu\nu}-\half (\pd .A)^2
\eqno(2.1)
$$
The canonical momenta conjugate to the space components $A^i$ of the
gauge field are
$$
\pi ^i=F^{0i}
,\eqno(2.2a)
$$
while that conjugate to $A^0$ is
$$
\pi ^0=-\pd .A
\,.\eqno(2.2b)
$$
The Hamiltonian density is
$$
H=\pi ^0\dot A^0+\pi ^i\dot A^i - {\cal L}
 =H_0 + H^{\hbox{{\sevenrm INT}}}
{}.\eqno(2.3)
$$
Here, $H_0$ is the part of $H$, expressed as a function of the fields
and the canonical momenta with no explicit time derivatives of the
fields, that survives if the coupling $g$ is set equal to zero, and
$$
H^{\hbox{{\sevenrm INT}}}(\pi ,A)=-g\pi ^i.A^0\wedge A^i
  +\quarter (F^{ij}F^{ij})^{\hbox{{\sevenrm INT}}}
{}.\eqno(2.4)
$$
To derive perturbation theory, one introduces
interaction-picture operators $Q_I(t)=\Lambda (t)Q(t)\Lambda ^{-1}(t)$,
where $\Lambda (t)=e^{iH_{0I}(t-t_0)}e^{-iH(t-t_0)}$.
These operators evolve with time as free fields: they obey Hamilton's
equations of motion with Hamiltonian $H_{0I}$ and the relation between
the interaction-picture canonical momenta and the fields is as in (2.2)
but with the coupling $g$ set equal to 0:
$$
\pi ^i_I=\dot A^i_I-\pd ^i A^0_I,\phantom{ZZZZZZZZZZZZ}
\pi ^0_I=-\pd .A_I.
\eqno(2.2c)
$$
The interaction-picture field equations are just the zero-mass
Klein-Gordon equations. Their solutions are
$$
A_I^{\mu}(x)=\int{{d^3k}\over{(2\pi)^3}}{1\over{2|{\k |}}}e^{-ik.x}
a^{\mu}(k) +\hbox{h c}
\eqno(2.5)
$$
where $k^0=|{\k}|$.
The canonical equal-time commutation relations
$[A_I^{\mu}(t,{\bf x}),\pi _I^{\nu}(t,{\bf y})]=i\d ^{\mu\nu}\d ^{(3)}
({\bf x}-{\bf y})$ etc give
$$
[a^{\mu}(k),a^{\nu\dag}(k')]=-g^{\mu\nu}2|{\k}|
(2\pi )^3\d ^{(3)}(\k -\k ')
,$$
$$
[a^{\mu}(k),a^{\nu}(k')]=[a^{\mu\dag}(k),a^{\nu\dag}(k')]=0
.\eqno(2.6)
$$

We now have to identify the physical states. For finite values of $t_0$,
the time at which the interaction picture coincides with the Heisenberg
picture, the constraint that they satisfy is non-trivial\ref{4}. However,
to derive zero-temperature perturbation theory, or finite-temperature field
theory in the real-time formalism, one may switch off the gauge coupling
adiabatically and work with the limit $t_0\to -\infty$. Then the
interaction-picture states are just the $in$-states and the condition that
picks out the physical states is the same as in abelian gauge theory.
Its most general form is that matrix elements of $\pd .A_I$ vanish between
physical states, but we may impose the stronger condition\ref{3}
$$
a^0(k)\;|\;\hbox{phys},\; in\;\rangle=0=\k .{\bf a}(k)\;|\;
\hbox{phys},\; in\;\rangle.
\eqno(2.7)
$$
In particular, the vacuum is required to satisfy these conditions.
Together, (2.5), (2.6) and (2.7) make the vacuum expectation
values of $T$-products of the fields take the usual  Feynman-gauge-propagator
forms.

The time-development operator for the interaction-picture states
away from $t=t_0$ is $\Lambda (t)$. By converting the differential
equation it satisfies into an integral equation and iterating,
one finds as usual that the $S$-matrix is a time-ordered exponential
of an integral of $H^{\hbox{{\sevenrm INT}}}(A_I,\pi _I)$.
$H^{\hbox{{\sevenrm INT}}}$ is given in (2.4). At this stage
we may replace the $\pi _I$ with their expressions (2.2c) in
terms of $A _I$ and $\dot A _I$. This gives
$$
H^{\hbox{{\sevenrm INT}}}=
  -{\cal L}^{\hbox{{\sevenrm INT}}}(A_I,\dot A _I)
  +\half g^2 (A_I^0 \wedge A_I^i)^2
.\eqno(2.8)
$$
The last term here compensates  for the presence of
$g\dot A^i_I.A_I^0\wedge A_I^i$ in the interaction Lagrangian\ref{5}.
The Hamiltonian formalism of perturbation theory that we have
outlined requires that, when the $\dot A_I^i$ field propagates
between two such neighbouring vertices one should use the propagator
$\langle 0|T\dot A_I^i(x_1)\dot A_I^j(x_2)|0\rangle$. However, the usual
Feynman rules use instead the double time derivative of the $A_I$-field
propagator, where the two time
differentiations are applied also to the $\theta$-functions that
appear in the definition of the $T$-product. The last term in (2.8)
is then simply omitted\ref{5}.

{}From the differential equation satisfied by $\Lambda (t)$ it is
straightforward to show that the $S$-matrix is unitary. However,
this does not imply that probability is conserved, because the
completeness relation includes the unphysical states. In order to
achieve probability conservation one must add to the Lagrangian
(2.1) a ghost part. This then guarantees that physical initial states
do not scatter into unphysical final states. One may show this
directly in the operator formalism\ref{4}, though we shall be content here
to accept the usual Faddeev-Popov path integral arguments as
justification.

We now extend the perturbation theory to nonzero temperature.
The same operator $\Lambda$ enters again, now with complex argument.
Simple algebra gives\ref{6}
$$
Z=\sum _i \;\langle \;i\;|e^{-\b H_{0I}}\L (t_0-i\b )\L ^{-1}(t_0)|\;i\;\rangle
$$
$$
\langle Q(t) \rangle =Z^{-1}\sum _i \;\langle \;i\;|e^{-\b H_{0I}}\L (t_0-i\b )
\L ^{-1}(t)Q_I(t)\L (t)\L ^{-1}(t_0)|\;i\;\rangle
\eqno(2.9)
$$
where $t_0$ must be the time at which the interaction picture coincides
with the Heisenberg picture and the sum is still over all physical
states, which we now choose to be interaction-picture $in$-states.

In order now to derive the Feynman rules, it is necessary first
to check that Wick's theorem may be used as usually in thermal field
theory. For this, note that the physical-gluon operators commute
with the unphysical gluon and ghost fields. Also, $H_{0I}$ is a sum of
commuting parts containing respectively only
physical and unphysical operators, $H_{0I}=H^{\hbox{{\sevenrm PHYS}}}_{0I}
+H^{\hbox{\sevenrm {UNPHYS}}}_{0I}$, and $H^{\hbox{\sevenrm {UNPHYS}}}_{0I}$
has zero eigenvalue in the physical states.
Hence any unphysical operators in the product
of operators in (2.9) may be factorised out:
only the vacuum expectation value of their product
appears, and the normal zero-temperature Wick's theorem applies to this.
For the
remaining  factor, we may replace $e^{-\b H_{0I}}$ by
$e^{-\b H^{ \hbox{{\fiverm PHYS}}}_{0I}}$;
the normal finite-temperature Wick's theorem applies to
this factor.

So the real-time finite-temperature perturbation theory is much as
usual\ref{2}, except that for the unphysical fields the propagators
are just vacuum expectation values (1.2), both for the unphysical
components of $A$ and for the ghost.

The gauge-field propagator in Feynman gauge therefore reads
$$\el{
{\bf D}_\mn =& -g_\mn
\(\matrix{\D_F & e^{\s k_0}\th(-k_0)\(\D_F-\D_F^*\)\cr
e^{-\s k_0}\th(k_0)\(\D_F-\D_F^*\) &-\D_F^*\cr} \)\cr
&+T_\mn\(\D_F-\D_F^*\){1\0e^{\b|k_0|}-1}
\(\matrix{1&e^{\s k_0}\cr e^{-\s k_0}&1\cr}\),\cr}
\en(2.10)
$$
with $0\le\s\le\b$, and $\D_F=1/(k^2+i\e)$.
The choice $\s=\b/2$ is usually made on grounds that this
makes the real-time propagator a symmetric $2\times2$ matrix.
Here this is the case only for the transverse projection
$T^{\m\r}{\bf D}_{\r\s}T^{\s\n}$.
A more natural choice is $\s=0$,
for it renders the vacuum piece independent of $\b$.
In the Appendix, we discuss various options for the diagonalisation
of (2.10).

Although the derivation was done in Feynman gauge, it seems obvious
how the resulting propagator will read in other gauges.
In a general linear gauge with quadratic gauge breaking term
${1\02\xi}(A^\m f_\m f_\n A^\n )$, where $f_\mu$ is a
4-vector which is either constant or constructed from derivatives
${\pd / \pd x}$,
the vacuum piece generalises by
$$g_\mn\to
G_\mn(k)=g_\mn-{k_\m f_\n+f_\m k_\n\0f.k}
+(f^2-\xi k^2){k_\m k_\n\0(f.k)^2}
.\en(2.11)
$$
Here $f=f(ik)$, if the gauge breaking term contains derivatives.
With $f_\m=ik_\m$, this reproduces the propagator in covariant
gauges, while $f_\m=(0,i\k )$ corresponds to the Coulomb gauge.
The ghost propagator is determined by $\D^{-1}_{\rm ghost}=f.k$;
its $2\times2$ matrix structure is analogous to the vacuum
part of the gauge propagator.
In most gauges, other than Feynman or Coulomb, an infinitesimal
nonhermitian piece has to be included in $G_\mn$
in order to give meaning
to the denominators in (2.11). In this case, the vacuum part of ${\bf D}_\mn$
becomes
$$
-\(\matrix{\D_FG_\mn  & e^{\s k_0}\th(-k_0)\(\D_FG_\mn -\D_F^*G_\mn ^* \)\cr
e^{-\s k_0}\th(k_0)\(\D_FG_\mn -\D_F^*G_\mn ^* \) &-\D_F^*G_\mn ^*\cr} \)
.\eqno(2.12)
$$
Although, strictly speaking, we have given a derivation only for the
Feynman gauge, the results we have quoted agree with those derived by James
for the $A_0=0$ gauge\ref{7}.
For general covariant gauges, they agree with
those in the literature\ref{8} if one sets the temperature to zero in the
non-$T^\mn$ part of the propagator.

\bigskip
3. {\bf CALCULATIONS}
\medskip
3.1. {\bf One-loop gluon self-energy}
\medskip
In a high-temperature expansion, the leading-temperature contributions
arise from one-loop diagrams, and they determine the physics at the
scale $gT$, where $g$ is the coupling constant.
In particular, the one-loop self-energy determines the spectrum
of quasiparticles in the plasma, and contains the information
on 
screening and Landau damping. 
It has been shown by explicit calculations\ref{9} and also
by an analysis of the relevant Ward identities\ref{10} that
these ``hard thermal loops'', as they have been called, are
independent of the gauge conditions needed to define the
gluon propagator. With the usual calculational procedure,
this gauge independence arises in
a rather nontrivial manner, and in particular in covariant
gauges the contributions from the thermalised Faddeev-Popov
ghosts are decisive.
With our prescription, 
the latter are absent
and it turns out to be
much simpler to establish the gauge independence.

With the usual prescription for covariant gauges,
the Feynman diagrams containing the
hard-thermal-loop contributions to the gluon self-energy
are those given in figure~1, where a slashed line denotes the thermal
part of a propagator, and a bare line the vacuum part.
With both vertices of type 1, these diagrams give Re~$\Pi_{11}$
which coincides with the real part of the
proper self-energy correction $\Pi$
for timelike momenta. 

In our formalism, we do not have contributions
from the ghost diagrams 1c, 1d,
although in diagram~1a the gauge-dependent
vacuum gluon propagator does contribute. We shall show that the
result is nevertheless gauge independent.

With our real-time Feynman rules,
the diagram~1b is identical to the corresponding
one in Coulomb gauge. Since in Coulomb gauge there are also no
contributions from the ghost diagrams,
diagram~1a has to be shown to coincide with the corresponding one
in the Coulomb gauge.
Evaluating diagram~1a in Coulomb gauge, one
finds for its leading-temperature part in SU(N) gauge theory
$$
\P^{(a){\rm CG}}_\mn(q)=
g^2N\int{d^Dk\0(2\p)^{(D-1)}}\d(k^2)n(|k_0|)
\[ -(D-2){4k_\m k_\n
-2k_\m q_\n-2q_\m k_\n \0(k-q)^2}-{k_0^2\0(\9k-\9q)^2}
T_\mn(k) \], \en(3.1)$$
where
the principal value of the pole of the integrand is understood.
Apart from the piece containing the Feynman denominator,
only the terms with highest degree in $k$
contribute. For this reason, one may further simplify
$k_0^2/(\9k-\9q)^2\to1$ in the last term of (3.1).

On the other hand, with the Feynman-gauge propagator (2.10)
one obtains
$$
\P^{(a)}_\mn(q)=
g^2N\int{d^Dk\0(2\p)^{(D-1)}}\d(k^2)n(|k_0|)
\[ -(D-2){4k_\m k_\n
-2k_\m q_\n-2q_\m k_\n \0(k-q)^2}+{(k+q)^2\0(k-q)^2}
T_\mn(k) \]. \en(3.2)$$
Because of the presence of the delta-function $\d(k^2)$, the
numerator in the
last term may be rewritten as
$(k+q)^2\to-(k-q)^2+2q^2$, and the $q^2$ dropped because it does
not contribute to the hard thermal part. Thus (3.2) gives the
same leading-temperature terms as (3.1) does.

Adding the contribution from diagram 1b gives the well-known result
\ref{9}
$$
\P_\mn(q)=
g^2N(D-2)\int{d^Dk\0(2\p)^{(D-1)}}\d(k^2)n(|k_0|)
\[ g_\mn -{4k_\m k_\n
-2k_\m q_\n-2q_\m k_\n\0(k-q)^2} \]. \en(3.3)$$

In gauges other than the Feynman gauge, the vacuum part of the
gluon propagator has to be changed according to
(2.11). Diagram 1b is not affected by this, but diagram 1a receives an
additional contribution
$$\el{
g^2N\int{d^Dk\0(2\p)^{(D-1)}}&\d(k^2)n(|k_0|)
T^{\r\s}(k)\[ {2f^\b\0f.(k-q)}
+{(\xi(k-q)^2-f^2)(k-q)^\b\0[f.(k-q)]^2} \]\cr
&\times(k-q)^\a V_{\a\m\s}(k-q,q,-k)
V_{\r\n\b}(k,-q,q-k),\cr}\en(3.4)$$
where in general $f_\m=f_\m(k-q)$.
$V$ is the 3-gluon vertex (without the SU(N) colour factor),
which obeys
$$(k-q)^\a V_{\a\m\s}(k-q,q,-k)=g_{\s\m}(k^2-q^2)-k_\s k_\m
+q_\s q_\m. \en(3.5)$$
When this is inserted into (3.4), only the two terms quadratic in the loop
momentum $k$ contribute to the leading-temperature part, but
both get cancelled, the $k^2$ by the $\d$-function, and the
$k_\s k_\m$ by the transverse projector $T_{\r\s}(k)$.
Therefore the result (3.3) is completely independent of the
gauge fixing.

For completeness, we also give the result, in our formalism,
for the one-loop contribution to the damping of long-wavelength
transverse gluons. In the conventional formalism, the imaginary part of
the gluon self-energy
at one-loop order is given by the diagrams of figure~2 and is
strongly gauge
dependent. In covariant gauges, the result comes with an
unphysical sign, which was even taken to be suggestive of a
fundamental instability of the perturbative thermal ground state
(see reference 11 for a review).
In our formalism, only diagram 2a contributes, and
the result is obviously gauge independent; it
reads
$${\rm Im}(T_\mn \P^\mn_{11})(k_0,\90)=-{1\06\p}Ng^2T^2.\en(3.6)$$
By virtue of
Im$(T_\mn\Pi^\mn)={\rm tanh}(k_0\b/2){\rm Im}(T_\mn\Pi^\mn_{11})$
this gives a contribution to the damping rate of long-wavelength
gluons
$$\g^{\rm one-loop}={g^2TN\024\p}.\en(3.7)$$
This result coincides with the Coulomb-gauge
one of the conventional formalism\ref{11}, where also only transverse
gluons contribute to the imaginary part.
It is well-known by now\ref{12} that one-loop results for the
gluonic damping rate are incomplete, and the gauge independence of
our result (3.7) should not lead one astray.
Still, we find it gratifying that in our formalism the unphysical
modes are not able to contribute to (3.7), whereas normally in the bare
one-loop calculation they even give rise to a negative sign.

\bigskip
3.2. {\bf Two-loop gluon-interaction pressure}
\medskip
The thermodynamic pressure of the gluon plasma, $P=1/(\b V)\ln Z$,
is a physical quantity and therefore gauge independent. In the
following we shall calculate it up to two-loop order in order to
perform a further check on our Feynman rules. The diagrammatic
rules for calculating the partition function in the real-time
formalism have been described in reference~13, to which we refer
for more details.

At one-loop order, our formalism leads to exactly the same
expression for $P$ as the conventional one does in Coulomb gauge.
Differences occur starting at two-loop order. The diagrams to
be considered are given in figure~3, where thermal contributions
have to be inserted in all possible ways, keeping at least one of the
vertices appearing therein of type 1 \ref{13}.

Diagram 3b potentially gives rise to integrals involving three
powers of the Bose distribution function, but these we find
to vanish after performing the momentum algebra because of
the three delta functions associated with the distribution
functions, exactly as in the conventional formalism.

Next consider the diagrams where two lines are thermal.
Diagram 3a then identically reproduces the result of conventional
Coulomb gauge,
$$\el{
P_2^{(a)}=&
-\8{1\08}{N(N^2-1)g^2}\int
{d^4q\,d^4k\0(2\pi)^6}n(|q_0|)\d(q^2)n(|k_0|)\d(k^2)
2\(3-z^2\)\cr=&
-{1\0216}N(N^2-1)g^2T^4,\cr} \en(3.8)$$
where we introduced $z\equiv \9k.\9q/|\9k||\9q|$.
Diagram 3b with two thermal and one vacuum gluon propagator 
is different, though.
The vacuum piece of the gluon propagator contributes
potential gauge dependences. However,
because of (2.11), all gauge-dependent
terms have one 4-momentum contracted with a 3-gluon vertex,
and give rise to integrals of the form
$$\int d^4q\,d^4k\,n(|q_0|)\d(q^2)n(|k_0|)\d(k^2)
T^\mn(k)T^{\s\r}(q)
(k-q)^\a V_{\m\a\s}(-k,k-q,q)\cdots. \en(3.9)$$
Because of
$$(k-q)^\a V_{\m\a\s}(-k,k-q,q)=g_{\s\m}(q^2-k^2)-q_\s q_\m
+k_\s k_\m  \en(3.10)$$
it is easy to see that all the integrals of the form (3.9)
vanish --- $q^2$ and $k^2$ in (3.10)
disappear because the $\d$-functions in (3.9), and
the other terms also owing to the presence of the transverse projectors
$T^\mn(k)T^{\s\r}(q)$.
Thus the contributions from figure 3b which involve two powers of
distribution functions are completely gauge independent.
There are actually two such diagrams. The one with both 3-vertices
of type 1 gives
$$\el{
P_2^{(b)(11)}=
-\8{1\04}{N(N^2-1)g^2}\int&
{d^4q\,d^4k\0(2\pi)^6}n(|q_0|)\d(q^2)n(|k_0|)\d(k^2)\cr&\times
{-(1+z^2)(k+q)^2+8(1-z^2)(\9k^2+\9q^2)\0(k-q)^2},\cr}  \en(3.11a)$$
where the first term can be simplified according to
$(k+q)^2/(k-q)^2\to-1$  because of the delta functions, and
the last term vanishes after performing the integrals, making
this contribution
coincide with the one of conventional Coulomb gauge. This yields
$$P_2^{(b)(11)}=-{1\0432}N(N^2-1)g^2T^4. \en(3.11b)$$
Because the vacuum part of the gluon propagator (2.10)
does have a (12)-component, there is a second diagram of
the form 3b, where one vertex is of type 1, and the other
of type 2.
We find after carrying out the momentum algebra
that the latter vanishes
because of the presence of three $\d$-functions, $\d(q^2)
\d(k^2)\d((k-q)^2)$.

Finally, there are the diagrams with only one gluon propagator
thermal. Those diagrams, however, contain a
zero-temperature one-loop subgraph with its external lines
put on-shell by the thermal propagator, and so are removed by
renormalisation. The additional
diagrams containing both a type-1 and a type-2
vertex do not complicate this picture, for they turn out to
vanish identically.

Discarding the purely zero-temperature contributions, we have therefore that
the
gluon-interaction pressure at two-loop order is
given by the sum of (3.8) and (3.11),
$$P_2=-{1\0144}N(N^2-1)g^2T^4,\en(3.12)$$
reproducing the standard result\ref{13-15}.
Notice that the ghost diagram 3c did not contribute to (3.12),
although we worked in a general linear gauge.

\bigskip
4. {\bf CONCLUSION}
\medskip
We have demonstrated that in the real-time version of
finite-temperature perturbation theory
it is quite natural to include a thermal part in the gauge
propagator only for the two physical degrees of freedom
of the gluons, leaving the
unphysical gauge-field components and the Faddeev-Popov ghosts unheated.
This allows us to combine the simplicity of covariant gauges
for the zero-temperature part
with the advantages of the noncovariant ghost-free
gauges for the thermal contributions. For the thermal contributions we
single out the rest frame of the heat bath
and do not cause additional violation
of Lorentz symmetry.
Our derivation has been carried through for the Feynman gauge.
The generalization to arbitrary gauges is immediate, though there may be
complications from the prescriptions necessary to define the poles at
$f.k=0$ in (2.11).

We have successfully
tested these new Feynman rules in one- and two-loop
calculations, and in particular
have been able to demonstrate gauge independence
of the high-temperature part of the gluon self-energy
and for the 2-loop gluon-interaction pressure in a
remarkably simple (albeit nontrivial) and general manner.
In the conventional formalism, gauge independence of the hard-thermal
gluon self-energy has been explicitly checked only in certain classes
of gauges\ref{9}, whereas the 2-loop pressure has been
calculated so far only in Feynman\ref{13,14}
and axial\ref{15,16} gauges.

We have worked throughout in the real-time formalism. This enables us to
develop the perturbation theory in terms of $in$ fields, with the great
advantage that not only do the fields obey free-field equations, but also
we may ignore the interaction in the condition that picks out the physical
states.  We have not been able
to translate our prescriptions satisfactorily into an imaginary-time
formalism. In order to set up such a formalism, one would need to choose
an interaction picture that coincides with the Heisenberg picture at finite
time $t_0$. The condition that picks out the physical interaction-picture
states is then much more complicated\ref{4}. Even if we were able to implement
it, the fact that the unphysical-field propagators are temperature-independent
and so not periodic in imaginary time would lead to extra complications\ref{6}
and make the formalism difficult to use.

An interesting question, to which we hope to return in
future work, is how the present approach could be extended
to accommodate a resummation of finite-temperature perturbation
theory, as is mandatory for exploring physics at the
scale $g^2T$ in high-temperature quantum chromodynamics\ref{12}
(and, incidentially, for handling pinch singularities\ref{17}).
An important difference between the bare and the resummed
theory is that, because of the appearance of an additional
collective mode at the scale $gT$, (1.3) no longer covers
all of the physical modes in the plasma. Including the latter
in a formalism analogous to the one presented here might
prove to be of conceptual interest for
the resummation program\ref{12}, and might perhaps allow
an easier verification of its gauge independence.

\bigskip
{\sl One of us (PVL) is grateful to Peter van Nieuwenhuizen for patient
discussions of the Gupta-Bleuler formalism; AR wishes to thank
Patrick Aurenche and Randy Kobes for useful remarks.}

\bigskip
{\bf APPENDIX: Diagonalisation of the matrix propagator}
\medskip
The real-time version of finite-temperature field theory
can be reformulated by diagonalisation of the 2$\times$2
matrix propagators, which allows one to associate all thermal
contributions with the vertices.

A well-known possibility\ref{2, 13} is to diagonalise to a matrix constructed
from the ordinary Feynman propagator $\D_F$,
$$ \9{\tilde D}=\(\matrix{\D_F&0\cr
0&-\D_F^*\cr}\). \en(A.1)$$
In our formalism, the Feynman-gauge propagator may be written as
$$
{\bf D}^\mn=T^\mn\;{\bf M_T\tilde DM_T}
-(g^\mn+T^\mn)\;{\bf M_0\tilde DM_0}
\eqno(A.2a)
$$
with (when $\s=0$)
$$\9{M_T}=\sqrt{n(|k_0|)}\(\matrix{
e^{\b|k_0|/2}& e^{-\b k_0/2} \cr
e^{\b k_0 /2}& e^{\b|k_0|/2} \cr}\), \en(A.2b)$$
and $\9{M_0}$ the zero-temperature limit of $\9{M_T}$:
$$\9{M_0}=\(\matrix{
1            & \th(-k_0)     \cr
\th(k_0)     & 1             \cr}\). \en(A.2c)$$
The ghost propagator is ${\bf M_0\tilde DM_0}$.

Recently, a different scheme has been proposed
by Aurenche and Becherrawy\ref{18}, which diagonalises
to a matrix composed of
retarded and
advanced propagators,
$$
{\bf \bar D}=
        \(\matrix{\Delta _R  &  0 \cr
                  0  &  \Delta _A \cr}\),
\eqno(A.3)
$$
instead of Feynman propagators.
For this, one has to
introduce different matrices to be associated with
incoming and outgoing lines\ref{18}. In our formalism for the Feynman gauge
$$
{\bf D}^\mn=T^\mn\;{\bf U_T\bar DV_T}
-(g^\mn+T^\mn)\;{\bf U_0\bar DV_0}
\eqno(A.4a)
$$
with
$$\el{
\9{U_T} = \(\matrix{1  &  -n(k_0) \cr
                  1  &  -(1+n(k_0)) \cr}\),
&\quad
\9{U_0} = \(\matrix{1  &\th(-k_0) \cr
                  1  &-\th(k_0) \cr}\),
\cr
\9{V_T} = \(\matrix{1+n(k_0)  &  n(k_0) \cr
                  1  &  1 \cr}\),
&\quad
\9{V_0} = \(\matrix{\th(k_0)  &-\th(-k_0) \cr
                  1  & 1 \cr}\).
\cr}      \en(A.4b)$$
The ghost propagator is ${\bf U_0\bar DV_0}$.
The matrices {\bf U} and {\bf V} are removed from the propagators
and instead associated
with the in- and outgoing lines of the vertices.
(Because of
$$
{\bf U}(k)\tau_1={\bf V}^T(-k),\quad \tau_1=\(\matrix{0&1\cr1&0\cr}\),
$$
more symmetric Feynman rules result if a factor
$\tau_1$ is combined with $\9U$ and
$\9{\5D}$, turning the latter off-diagonal\ref{19}.)
With (A.4),
the analysis of reference~18 carries over to a large extent.
However, in addition to retarded and advanced lines one has
to distinguish between transverse and non-transverse ones,
and in particular there is no simple causality principle
when transverse and non-transverse lines come together at
a vertex. For example, a 3-vertex connecting solely
retarded or solely advanced lines does not vanish, unless
the lines are either all transverse or all non-transverse.

Such a reformulation in terms of retarded and advanced Green functions
leads to well-defined causal properties, which are not manifest
in the usual real-time formalism\ref{20}.
Causal Green functions, which are relevant for example in
linear response theory, are usually obtained in a direct manner
in the imaginary-time formalism. In our prescription,
however, this is not straightforward to implement, and so
this diagonalization scheme
appears to be particularly useful for our approach.

\vfill\eject
{\bf REFERENCES}
\medskip
\i{1} C W Bernard, Phys Rev D9 (1974) 3312; H Hata and T Kugo, Phys Rev
D21 (1980) 3333

\i{2} L V Keldysh, Soviet Physics JETP 20 (1965) 1018;
A Niemi and G W Semenoff, Nucl Phys B230 (1984) 181

\i{3} C Itzykson and J-B Zuber, {\sl Quantum field theory} (McGraw-Hill,
1980) p 127

\i{4} T Kugo and I Ojima, Phys Lett 73B (1978) 459 and Prog Theor Phys
60 (1978) 1869; 61 (1979) 294; 61 (1979) 644

\i{5} P T Matthews, Phys Rev 76 (1949) 1657; P V Landshoff and J C Taylor,
Phys Lett B231 (1989) 129

\i{6} K A James and P V Landshoff, Phys Lett B251 (1990) 167

\i{7} K A James, Z Phys C48 (1990) 169.

\i{8} R L Kobes, G W Semenoff and N Weiss, Z Phys C29 (1985) 371;
N P Landsman, Phys Lett B172 (1986) 46;
K L Kowalski, Z Phys C36 (1987) 665

\i{9} O K Kalashnikov and V V Klimov, Sov J Nucl Phys 31 (1980) 699;
H A Weldon, Phys Rev D26 (1982) 1394;
J Frenkel and J C Taylor, Nucl Phys B334 (1990) 199;
E Braaten and R D Pisarski, Nucl Phys B337 (1990) 569; B339 (1990) 310

\i{10} R Kobes, G Kunstatter and A Rebhan, Nucl Phys B355 (1991) 1

\i{11} U Heinz, K Kajantie and T Toimela, Ann Phys (NY) 176 (1987) 218

\i{12} R D Pisarski, Nucl Phys A525 (1991) 175c

\i{13} N P Landsman and Ch G van Weert, Phys Rep 145 (1987) 141

\i{14} J I Kapusta, Nucl Phys B148 (1979) 461

\i{15} K A James, Z Phys C49 (1991) 115

\i{16} H Nachbagauer, PhD thesis, Tech Univ Vienna (1991);
Z Phys C56 (1992) 407

\i{17} H A Weldon, Phys Rev D45 (1992) 352

\i{18} P Aurenche and T Becherrawy, Nucl Phys B379 (1992) 259

\i{19} M A van Eijck and Ch G van Weert, Phys Lett B278 (1992) 305

\i{20} R L Kobes, Phys Rev D42 (1990) 562; D43 (1991) 1269
\bigskip
\bigskip

{\bf Figure captions}
\medskip
\i{1} Diagrams which contain the hard-thermal-loop contributions
to the gluon self energy. Solid lines are
gluons and broken lines Faddeev-Popov ghosts. The slashes denote thermal
parts of the propagators; unslashed lines are vacuum propagators.
In the formalism described in this paper, diagrams
(b) and (c) do not occur.

\i{2} The imaginary part of the gluon self-energy,
with the same notation as
figure 1. In our formalism, (b) does not occur.

\i{3} Diagrams for the 2-loop gluon
interaction pressure. The lines represent complete
propagators (vacuum plus thermal parts).
\bye



\hsize=17truecm\hoffset=-5truemm
\vsize=25.5 truecm\voffset=-5truemm
\nopagenumbers
\input pictex.tex
 \font\lab=cmr10 at 12 truept
\linethickness=1truemm
\setcoordinatesystem units <0.8truemm, 0.8truemm> point at 0 0
\beginpicture

\newbox\picA
\setbox\picA=\hbox{\beginpicture
 \replot "gupta.a"
\endpicture}

 \lab
\put{\copy\picA} at 0 0
\putrule from 35 5 to 35 15
\put{(a)} at 35 -20

 \replot "gupta.c"
\putrule from 135 15 to 135 25
\put {+} at  90 0
\put {1} at 100 5
\put {2} at 100 -5
\put {---} at 100 0
\put {(b)} at 135 -20

\newbox\picB
\setbox\picB=\hbox{\beginpicture
 \replot "gupta.b"
\endpicture}

\put{\copy\picB} at 10 -55
\put {---} at 0 -55
\putrule from 45 -50 to 45 -40
\put {(c)} at  45 -80

\put{\copy\picB} at 110 -55
\put {---} at 95 -55
\putrule from 145 -70 to 145 -60
\put {(d)} at 145 -80

\put {Figure 1} at 95 -93

\put{\copy\picA} at 10 -135
\put {---} at 0 -135
\put {1} at 0 -130
\put {2} at 0 -140
\putrule from 45 -130 to 45 -120
\putrule from 45 -150 to 45 -140
\put {(a)} at 45 -160

\put{\copy\picB} at 110 -135
\put {---} at 97 -135
\putrule from 145 -130 to 145 -120
\putrule from 145 -150 to 145 -140
\put {(b)} at 145 -160
\put {Figure 2} at 95 -173

\replot "gupta.d"
\put {\lab ---} at 0 -215
\put {\lab ---} at 10 -215
\put {\lab 1} at 10 -210
\put {\lab 8} at 10 -220
\put {\lab (a)} at  37 -235

\put {\lab ---} at 69 -215
\put {\lab ---} at 77 -215
\put {\lab 1} at 77 -210
\put {\lab 12} at 77 -220
\put {\lab (b)} at 105 -235

\put {\lab +} at 135 -215
\put {\lab ---} at 142 -215
\put {\lab 1} at 142 -210
\put {\lab 2} at 142 -220
\put {\lab (c)} at 170 -235
\put {\lab Figure 3} at 95 -248
\endpicture

\bye